\documentclass[letterpaper,twocolumn,10pt]{article}
\usepackage{usenix2022_SOUPS}

\usepackage{tikz}
\usepackage{amsmath}

\usepackage{filecontents}
\usepackage{tabularx}
\usepackage{booktabs}
\usepackage{graphicx}
\usepackage{hyperref}
\usepackage{hhline}
\usepackage{subcaption}
\usepackage{enumitem}

\usepackage{float} 
\usepackage{soul,color}
\usepackage[utf8]{inputenc}
\usepackage{longtable}
\usepackage[flushleft]{threeparttable}
\usepackage{chngcntr}
\usepackage{xcolor}
\definecolor{editCol}{rgb}{0.0, 0.0, 0.0}
\newcommand{\edit}[1]{{\textcolor{editCol}{#1}}}
\usepackage{pbox}
\usepackage{multirow}

\usepackage{array}
\usepackage{makecell}
\usepackage{afterpage} 
\usepackage[title]{appendix}
\usepackage{subcaption,graphicx}
\usepackage{ragged2e}

\newcolumntype{L}[1]{>{\raggedright\let\newline\\\arraybackslash\hspace{0pt}}m{#1}}
\newcolumntype{C}[1]{>{\centering\let\newline\\\arraybackslash\hspace{0pt}}m{#1}}
\newcolumntype{R}[1]{>{\raggedleft\let\newline\\\arraybackslash\hspace{0pt}}m{#1}}

\begin{document}

\date{}

\title{\Large \bf Evaluating the Impact of Community Oversight for Managing Mobile \\Privacy and Security}

\def\plainauthor{Author name(s) for PDF metadata. Don't forget to anonymize for submission!}



\author{
{\rm Mamtaj Akter}\\
Vanderbilt University
\and
{\rm Madiha Tabassum}\\
Northeastern University
\and
{\rm Nazmus Sakib Miazi}\\
Northeastern University
\and
{\rm Leena Alghamdi}\\
University of Central Florida
\and
{\rm Jess Kropczynski}\\
University of Cincinnati
\and
{\rm Pamela J. Wisniewski}\\
Vanderbilt University
\and
{\rm Heather Lipford}\\
University of North Carolina, Charlotte
}

\maketitle
\thecopyright

\begin{abstract}

Mobile privacy and security can be a collaborative process where individuals seek advice and help from their trusted communities. To support such collective privacy and security management, we developed a mobile app for Community Oversight of Privacy and Security  ("CO-oPS") that allows community members to review one another's apps installed and permissions granted to provide feedback. We conducted a four-week-long field study with 22 communities (101 participants) of friends, families, or co-workers who installed the CO-oPS app on their phones. \edit{Measures of transparency, trust, and awareness of one another's mobile privacy and security behaviors, along with individual and community participation in mobile privacy and security co-management, increased from pre- to post-study.} Interview \edit{findings confirmed} that the app features supported collective considerations of apps and permissions. However, participants expressed a range of concerns regarding having community members with different levels of technical expertise and knowledge regarding mobile privacy and security that can impact motivation to participate and perform oversight. Our study demonstrates the potential and challenges of community oversight mechanisms to support communities to co-manage mobile privacy and security.

\end{abstract}

\section{Introduction}
The majority of U.S. adults own smartphones \cite{nw_demographics_nodate}, and nearly half of them have reported downloading various third-party apps \cite{nw_mobile_2015}. These mobile apps often require access to users' sensitive information, such as contacts, emails, location, photos, calendars, and even browser history \cite{nw_mobile_2015}. Most apps request users' permission before accessing any information or resources. Yet users may have difficulty understanding these permission requests and the implications of granting them ~\cite{ferreira_securacy_2015, park_privacy_2022, alsoubai_permission_2022}. As a result, users struggle to make permission decisions or grant permission by mistake \cite{kelley_conundrum_2012}. Even worse, there are ways for more malicious apps to circumvent the permissions system and secretly gather users' system resources and private information without consent \cite{reardon_50_2019}. Ironically, a recent Pew Research study reported that most US adults are concerned about how their personal information is being used by these third-party apps as respondents felt they lack control over their mobile privacy \cite{vogels_americans_2019,  davis_perceived_1989}. 

This lack of understanding leads users to seek advice and guidance from others ~\cite{dourish_security_2004}. Several studies have demonstrated that users often learn about privacy and security from their social network, which influences them to change their own digital privacy and security behavior ~\cite{schechter_learning_2015, felt_android_2012, mendel_soial_2023}. As such, networked privacy researchers acknowledged the importance of these social processes for managing individual and collective digital privacy and security ~\cite{das_effect_2014, mendel_susceptibility_2017, rader_identifying_2015}. Despite this prior work, few mechanisms to support these social processes have been developed and evaluated. In this paper, we explore community oversight, where trusted groups of users help one another manage mobile privacy and security. 
\textcolor{black}{In our previous work, we proposed a theoretical framework of community oversight \cite{chouhan_co-designing_2019}, describing how the concepts of transparency, awareness, trust, individual and community participation are needed within a particular mechanism.} We have now implemented a mobile app, Community Oversight of Privacy and Security (CO-oPS), to explore these concepts in use and support a collaborative approach to mobile privacy and security management. The CO-oPS app allows individuals in a community to review one another's apps installed and permissions granted and provide direct feedback to one another.

In this paper, we present a field study of the CO-oPS app. Our aim was to understand the impact of using the app on participants' mobile app decisions and perceptions. We conducted a 4-week mixed-method longitudinal field study with 101 people in 22 self-formed groups. Each group installed, used, and evaluated the CO-oPS app, provided oversight to one another on their mobile app privacy decisions, and shared experiences through weekly surveys and optional interviews. We describe how users interacted within the app and the changes in their mobile app permission decisions after using the CO-oPS app. We also examine how participants' perceptions regarding co-managing their mobile privacy and security within their communities change throughout the study. 
\textcolor{black}{To do so, we measured constructs derived from our community oversight model \cite{chouhan_co-designing_2019} of perceptions of transparency, awareness, trust, and individual and community participation within the CO-oPS app.} \edit{We tested for the pre-post study differences and detected increases for all of these measures that were statistically significant.} Qualitative findings further explain these perceptions and identify co-management concerns: feelings of privacy invasion of their own and others, lack of trust in less knowledgeable community members, lack of close relationships, and communities' inadequate tech expertise. We also found that using the CO-oPS app helped participants increase their communities' collective capacity to address their mobile privacy and security concerns.

In sum, our study makes a unique contribution to SOUPS research community by \edit{investigating through a field study} how a community oversight mechanism can help increase participants' collective capacity to support one another in co-managing mobile privacy and security together as a community. Specifically, we make the following unique research contributions: 1) Through a longitudinal field study, we describe the benefits and challenges of using a community oversight app to co-manage mobile privacy and security; 2) We provide empirical evidence of the potential for community oversight to increase users' awareness of mobile privacy issues, leading to individual changes in decisions and community exchange of knowledge; and 3) We present considerations and design-based recommendations towards features to support communities in providing oversight to one another.


\section{Background}

\noindent
\textbf{Privacy and Security Management in Mobile Applications}

Mobile applications often access sensitive information and share users' personal data with third parties ~\cite{calciati_automatically_2020, reardon_50_2019, hatamian_hard_2019, feichtner2020understanding, lu_demystifying_2020}. As such, substantial work has been done to investigate and support end users in managing mobile app privacy and security. Researchers have looked at the existing privacy awareness and management approaches (e.g., app privacy permission prompts, privacy policies, etc.) and found that such mechanisms often fail to provide users with awareness and knowledge of privacy and security risks ~\cite{till_characterization_2019, ferreira_securacy_2015, almuhimedi_your_2015, felt_android_2012, kelley_conundrum_2012}. \edit{Moreover, users often do not understand mobile app permission dialogues \cite{ferreira_securacy_2015} and are over-exposed to such requests \cite{till_characterization_2019}.} Researchers have proposed several technology-based solutions to increase awareness and limit potential risks associated with third-party mobile apps ~\cite{sadeghi_temporal_2018, lutaaya_rethinking_2018, peddinti_reducing_2019, alghamdi2022webprototype}. \edit{For example, Sadeghi et al. suggested evaluating the app permissions against risks and automatically grant/revoke permission on users' behalf \cite{sadeghi_temporal_2018}. Others proposed mechanisms to inform users about the app privacy risks, recommend secure choices, and nudge them to review/revise permissions ~\cite{zhu_mobile_2014,liu_follow_2016,almuhimedi_your_2015}. Others suggested tools to allow users to review data before sending it to the server, visualize data flow \cite{bahrini_2019}, and replace personal information with mock data without affecting app functionality \cite{lutaaya_rethinking_2018}.}



While this body of research has emphasized enhancements to technology to help individuals manage privacy and security while using mobile applications, none looked at how knowledge and influence from social groups help in individual privacy and security decision-making. Our research focuses on assessing and supporting these social processes involved in privacy and security management.\\

\noindent
\textbf{Community-based Approaches for Privacy and Security}

\edit{In general, research shows that people frequently take collaborative approaches to make privacy and security decisions \cite{nthala_2018, rader_identifying_2015}, and users often rely on social factors while making such decisions. Chin et al. discovered that smartphone users are more likely to consider social signals, such as reviews and ratings from other users, rather than privacy indicators regarding Android permissions when making app use decisions \cite{chin_2012}. Das et al. demonstrated that social factors (e.g., community adoption of security features) could increase individuals' security awareness and encourage them to adopt security features \cite{das_the_2015}. As such, researchers have proposed using social and community influence to assist individuals in making decisions about digital privacy and security \cite{mendel_susceptibility_2017, Goecks_2005, squicciarini2011cope}. Squicciarini et al. developed CoPE, a tool to support users in collaboratively managing their shared images in social network sites \cite{squicciarini2011cope}.}

\edit{Past research has also examined privacy management approaches involving one party performing oversight for another. Organizations adopt mobile device management (MDMs) systems to remotely control and secure the data stored in employees' mobile devices \cite{hayes_effective_2020}. Parents use adolescent online safety apps to monitor and protect teens by restricting their online behavior ~\cite{akter_from_2022, wisniewski_parental_2017, agha_strike_2023, ghosh_safety_2018}. The results from these studies suggest that a collaborative approach, rather than one-sided control, could benefit both parties and lead to more privacy-preserving outcomes. Finally, several studies leveraged crowdsourcing to use mass user data to support individual users in making improved mobile privacy and security decisions \cite{Lin_expectation_2012, ismail2017permit, rashidi_Android_2018, zhang_privacy_2016, liu_understanding_2019}. For instance, Ismail et al. utilized crowdsourcing to recommend permissions that can be disabled for enhanced privacy without sacrificing usability \cite{ismail2017permit}. However, these approaches showed little consideration for the trustworthiness of information from a random crowd. \textcolor{black}{On the other hand,} researchers found that users are more willing to adopt and share privacy advice from a trusted community \cite{redmiles2016think}, and they often communicate first with friends and family to learn about potential privacy and security threats and mitigation strategies \cite{das_effect_2014}.} 

\edit{In summary, our work builds upon the past literature in social cybersecurity, MDMs, parental control apps, and crowdsourcing to implement and evaluate a novel model of community-based oversight (i.e., self-selected groups) for mobile privacy and security through a large-scale field study. Since the network structure of oversight (e.g., individual for MDMs, many-to-one for crowdsourced recommendations, and unidirectional from parent to child for parental control) in these prior works is vastly different than ours, this new model of community oversight warrants deeper empirical investigation. 
\textcolor{black}{In ~\cite{chouhan_co-designing_2019}, we were the first to propose a novel framework of community oversight for helping people manage their mobile privacy and security together. Through a participatory design study, we} identified mechanisms that would allow users to support others in the community in making privacy and security decisions regarding mobile app permissions. 
\textcolor{black}{We also designed a prototype mobile app that allows users to collaborate and share information with people they know to help make mobile app permissions decisions \cite{aljallad_designing_2019}.} While this body of our prior studies provides a valuable basis for the design of community-oriented privacy and security management systems, they only present a theoretical view of users' preferences in community decision-making. \textcolor{black}{In contrast, this study} contributes to the literature by providing an in-situ evaluation of how trusted groups of people use and interact with different community-oriented features to collaboratively manage their mobile privacy and security.}

\begin{figure}[t]
\centering
\begin{subfigure}[t]{.204\textwidth}\centering
  \includegraphics[width=\columnwidth]{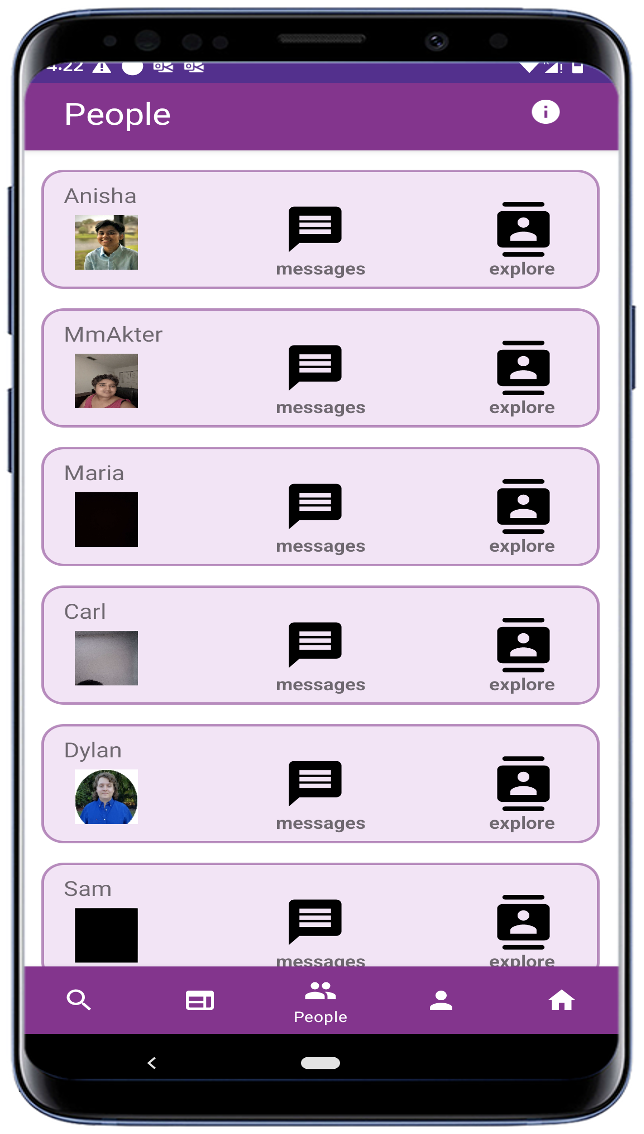}
  \caption{}
\end{subfigure}%
\begin{subfigure}[t]{.205\textwidth}\centering
  \includegraphics[width=\columnwidth]{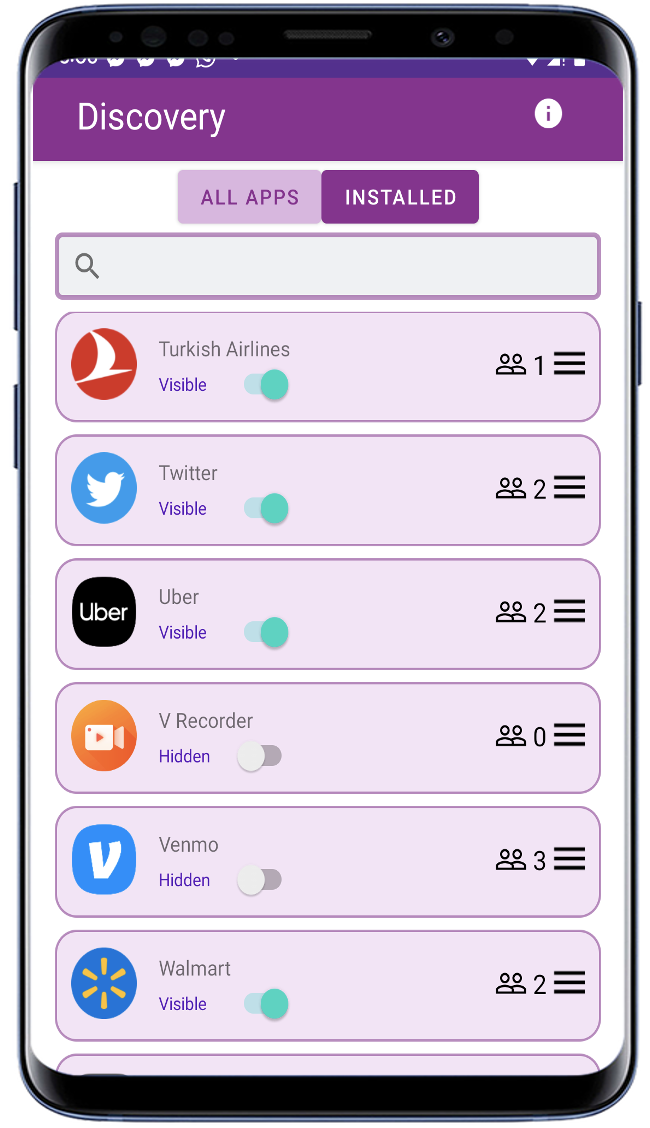}
  \caption{}
\end{subfigure}
\begin{subfigure}[t]{.205\textwidth}\centering
  \includegraphics[width=\columnwidth]{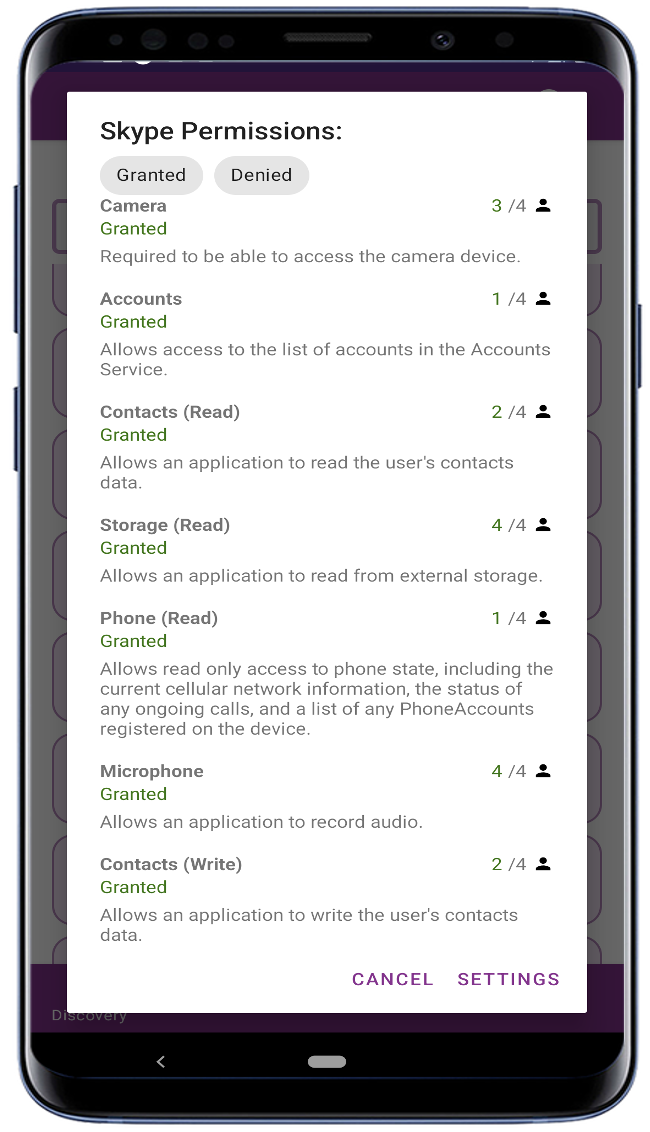}
  \caption{}
\end{subfigure}
\begin{subfigure}[t]{.206\textwidth}\centering
  \includegraphics[width=\columnwidth]{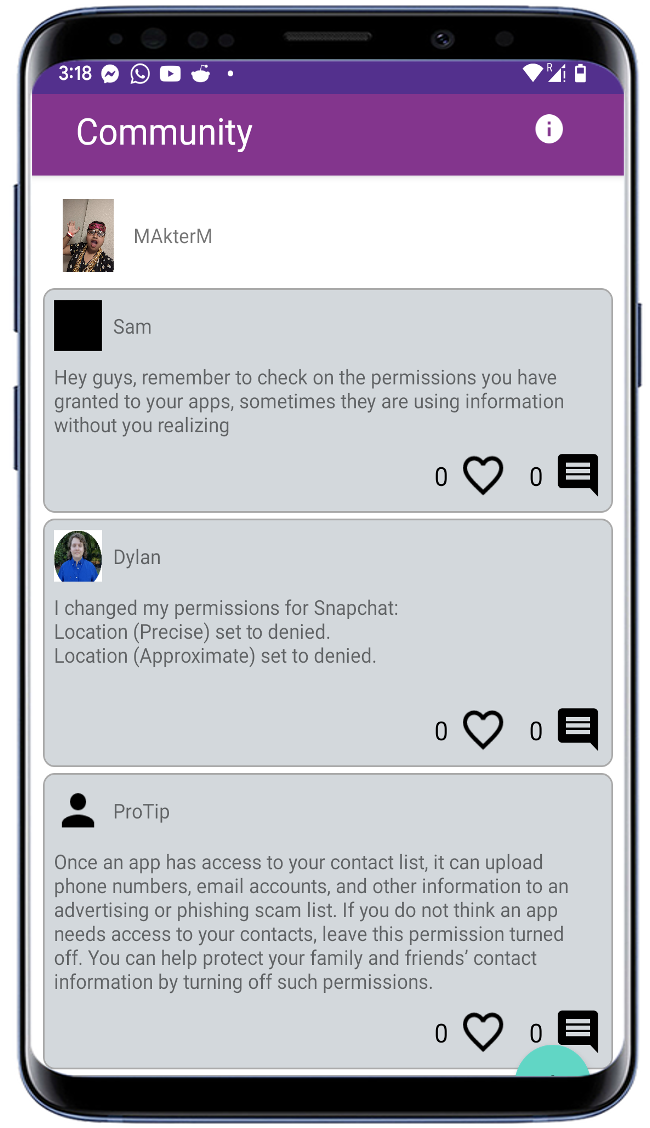}
  \caption{}
\end{subfigure}
\caption{CO-oPS Features: (a) People, (b) Discovery, (c) Permissions, (d) Community Feed.}~\label{fig:figure1}
\end{figure}

\section{Design of The CO-oPS App}
\textcolor{black}{We developed the Community Oversight of Privacy and Security (CO-oPS) Android app \cite{akter_CO-oPS_2022} based on the model of community oversight proposed in our prior work \cite{chouhan_co-designing_2019}}. This model outlines the need for community oversight mechanisms to support individual and community participation through awareness and transparency features that build trust between community members. Thus, our CO-oPS app design includes four key features: 1) People page, 2) Discovery, 3) Permissions, and 4) Community Feed. 
\edit{The Discovery page allows community members to review one another’s installed apps (Figure-\ref{fig:figure1}(b)), and the list of permissions granted or denied to each app (Figure-\ref{fig:figure1}(c)). Users also can review the count of total community members who have the same apps installed or permission granted. To help users change the app permissions easily, the Permission page provides a “SETTINGS” link that forwards users to Android Settings to modify app permissions. On the Discovery page, users can also hide some of their own apps from their community, ensuring their personal privacy. To provide feedback to one another, users can direct message and can openly discuss any privacy and security issues on the Community feed page (Figure-\ref{fig:figure1}(d)). This community feed has another important function: when someone in the community changes their app permission, the CO-oPS app creates an automatic post on the community feed about that change. It also posts weekly protips to educate community members regarding safe apps and permissions.}

\section{Study Constructs}
To evaluate the impact of using the CO-oPS app, we \edit{measured a set of constructs that we surveyed before, during, and at the end of the field study. We measured all constructs by presenting participants with various statements relevant to each construct. Participants were asked to rate each statement on a 5-point Likert scale from 1 (strongly disagree) to 5 (strongly agree). 
\textcolor{black}{First, we developed new constructs derived from the theoretical framework for community oversight proposed in our prior work \cite{chouhan_co-designing_2019}}, consisting of transparency, awareness, trust, individual participation, community participation, and community trust. We validated these new constructs through standard psychometric tests (i.e., Cronbach's alpha \cite{cronbach_coefficient_1951} to confirm internal consistency), which is reported in Table-\ref{tab:ttest}. Then, we utilized three pre-validated scales from prior research \edit{\cite{kropczynski_towards_2021, sarason1974psychological, carroll_collective_2005, carroll_community_2003}} to measure community belonging, self-efficacy, and community collective efficacy. All scale items are included in Appendix A.} Below, we define each of the constructs, along with our hypotheses.\\

\noindent\textbf{Transparency:}
As Das et al. demonstrated \cite{das_the_2015}, social proof - seeing others adopt a privacy and security behavior - often helps individuals adopt the same behavior. Therefore, to encourage individuals in a community to make informed decisions for their mobile privacy settings, the behaviors of others must first be transparent. Therefore we define transparency as an individual's perceived visibility of their community's mobile apps installed and the permissions granted/denied.

\noindent
\textit{\textbf{H1:} At the end of the study, community members will perceive higher levels of transparency in their community's mobile privacy and security behaviors. }\\

\noindent\textbf{Awareness:}
Endsley demonstrated \cite{endsley_toward_1995} that situational awareness - the understanding of what is going on around someone - is a key component in effective decision-making. In a later study \cite{dourish_social_2005}, DiGioia and Dourish suggested that being informed about digital privacy and security norms and practices along with the actions performed by the community are necessary for an effective social influence process. We developed our awareness measure as an individual's perception about the awareness of their own and others' apps installed, permissions granted/denied, along with the changes made.

\noindent
\textit{\textbf{H2:} At the end of the study, community members will perceive higher levels of awareness regarding their community's mobile privacy and security practices. }\\

\noindent\textbf{Trust:}
\textcolor{black}{In \cite{chouhan_co-designing_2019}, we identified that having the information available and being informed about mobile privacy and security practices might not be sufficient for community oversight.} This is because individuals need to be able to trust the quality of the information and perceive the information as dependable to learn from and be influenced by it. 

\noindent
\textit{\textbf{H3:} Community members will have a higher level of trust in one another's mobile privacy and security decisions.}\\

\noindent\textbf{Individual Participation:}
While an effective social process needs transparency, awareness, and trust in one another, individuals also need to be willing to engage in this process \cite{chouhan_co-designing_2019}. Users need to be motivated to utilize the knowledge gathered from their community in order to make decisions. They also need to be willing to provide oversight to others. Thus we define individual participation as an individual's willingness to take steps to make changes in their own mobile privacy and security behaviors (uninstalling unsafe apps or denying dangerous permissions) and also providing oversight to others' mobile privacy and security behaviors (providing feedback and guidance to others). 

\noindent
\textit{\textbf{H4:} Community members will perceive higher individual participation at the end of the study. }\\

\noindent\textbf{Community Participation:}
Community oversight mechanisms can take place in different types of communities, such as, families \cite{cranor_life_2014}, coworkers \cite{lipford_someone_2012}, friends, and social networks \cite{lipford_someone_2012}. Yet not all types of communities may have an equal level of willingness to take part in different forms of community oversight. 
\textcolor{black}{For example, in \cite{chouhan_co-designing_2019}, we found that communities with closer relationships might be more willing to help one another make decisions than communities with weaker ties.} Therefore, we define community participation as an individual's perception of their community to collectively work together, e.g., help one another, exchange feedback and guidance, and engage in open discussions.  

\noindent
\textit{\textbf{H5:} At the end of the study, participants will perceive a higher level of community participation. }\\

\noindent\textbf{Community Trust and Belonging:}
Individuals are likely to help one another if they feel like they belong and can trust their community members. We define \textit{Community Trust} as an individual's perception of trusting their community to keep their personal information (e.g., apps installed) private and care for one another's mobile privacy and security. For community belonging, we utilized a pre-validated measure \cite{carroll_community_2003, sarason1974psychological} that has been used in exploring community support mechanisms outside of privacy and security. The \textit{community belonging} construct measures an individuals' feelings about how much they matter to their community. While our participants already knew each other, participating together in the CO-oPS app could lead them to feel stronger bonds and care between each other. Therefore, our hypotheses are:

\noindent
\textit{\textbf{H6:} An individual's community trust will be higher at the end of the study.}

\noindent
\textit{\textbf{H7:} Community belonging will be higher after the study.}\\

\noindent\textbf{Efficacy:}
Two of the outcomes we wanted to measure are perceptions over the efficacy of individuals and groups to manage their mobile privacy and security. Thus, we used pre-validated measures for self-efficacy \cite{bandura1982self}, and community collective efficacy \cite{carroll_collective_2005} in our study. The \textit{self-efficacy} \cite{bandura1982self} construct measures an individual's perceived capacity to manage their own mobile privacy and security. The \textit{community collective efficacy} \cite{carroll_collective_2005, kropczynski_towards_2021} construct measures an individual's perceived collective capacity to manage their community's privacy and security together. Our hypotheses are:

\noindent
\textit{\textbf{H8:} Individual's self-efficacy will be higher after the study.}

\noindent
\textit{\textbf{H9:} Community collective efficacy will also be higher at the end of the study.}

\section{Methods}

\noindent
\textbf{Study Overview:} The overall goal of our study is to evaluate the CO-oPS app in building the capacity of the communities to manage their mobile privacy and security collectively. We also wanted to understand what impacts this community-based approach may have in changing participants' perceptions and behaviors toward their individual and collective mobile privacy and security management. To achieve these goals, we recruited small self-organized communities (2-6 Android phone users) who knew each other. Each community member installed the CO-oPS app and participated for four weeks. Measures were gathered before app installation, each week of the study, and at the end. Each week participants were asked to complete different in-app tasks that allowed them to explore the features of the CO-oPS app. Finally, participants were invited to participate in an optional follow-up interview. \edit{In each step of the study, we explicitly provided the definition of the term "community" as "your group members who are participating in this study."} Each participant was compensated with a \$40 Amazon gift card for completing the field study, with an additional \$10 Amazon gift card for participating in the interview. Some participants withdrew from the study after two weeks due to technical difficulties with their smartphones and were compensated half the amount. Twenty-nine participants discontinued participation after week one, \edit{perhaps due to natural attrition}, and were not compensated. Data were discarded from all who did not complete the study. \\

\begin{table}[hbt!]
 \footnotesize
 \centering
\caption{Sociodemographic Characteristics of Participants}
  \label{tab:demo}
\begin{tabular}{llcc}
\toprule

& Total no. of participants                & \textit{N=101} & \textit{100\%} \\ \hline
\hline
Gender   & Female                  & 46         & 45.5   \\
& Male                    & 55         & 54.5    \\ \hline
Age     & 13-17                   & 6          & 5.9     \\
& 18-24                   & 27         & 26.7   \\
& 25-34                   & 49          & 48.5    \\
& 35-44                   & 6          & 5.9   \\
& 45-54                   & 10          & 9.9    \\
& 55-64                   & 1          & 1    \\
& 65+                     & 2          & 2     \\
 \hline
Ethnicity      & Asian/Pacific Islander          & 72          & 71.3      \\
& Black/African American             & 13         & 12.8    \\
& Hispanic/Latino                 & 8         & 7.9  \\
& White/Caucasian                 & 8          & 7.9     \\ \hline

 Education               &   Primary School          & 8          & 7.9      \\
& High School             & 5         & 5     \\
& College (Associate)                 & 6         & 5.9  \\
& College (Bachelor)                 & 40         & 39.6  \\
& Masters                 & 36          & 35.6     \\
& Doctorate   & 6          & 5.9    \\ 
\bottomrule
\end{tabular}
\end{table}

\noindent
\textbf{Participant Recruitment: }
We recruited a total of 101 participants that were associated with 22 communities. We initially recruited the primary contacts of each community who completed a pre-screening eligibility survey that verified whether they met the inclusion criteria of the study prior to providing their informed consent. The inclusion criteria for participation included: 1) reside in the United States, 2) be 13 years or older, 3) have an Android smartphone, and 4) be willing to install and use the CO-oPS app. \edit{Here, we also specified that they “must participate in a group with two other people you know," which determined the minimum group size required to participate in this study. After completing the screening survey, the initial contacts} were asked to share this eligibility survey with people they knew to invite them to participate in this study as their community members. \edit{Therefore, the initial contact of each group self-selected their community based on the above criteria (1-4). As such, all group members knew the initial contact but in some cases, were only loosely acquainted with one another.} For the teen participants, we required their parents to complete this survey and provide their consent. 

Our study was Institutional Review Board approved. \edit{The target characteristics of our participants were all Android smartphone users of any age range (minors, adults, and older adults). Therefore, we did widespread recruitment through social media, email, phone calls, and word-of-mouth.} The recruitment process started in January 2022 and ended in August 2022. Overall, we recruited 22 communities (101 participants) where the size of the communities ranged from 2 to 6. Table-\ref{tab:demo} summarizes the gender, age groups, ethnicity, and education of our participants. Our participants were primarily young, between the ages of 13 to 34. Most of them had a college degree. The majority of the participants were Asian, followed by African American, Hispanic/Latino, and White/Caucasian. \edit{Table-\ref{tab:chargroup} illustrates the frequency of the group compositions. Most of the groups consisted of family members, friends, and others (e.g., neighbors, co-workers, and acquaintances). }

\begin{table}[tbh]
 \footnotesize
\centering
 \caption{\edit{Group Compositions}}
  \label{tab:chargroup}
\begin{tabular}{llrllr}
\toprule

Total no. of groups                   & \textit{22} & \textit{100\%} \\\hline
 \hspace{3mm}Family Only                         & 2           & 9.1           \\
 \hspace{3mm}Family and Friends                          & 4           & 18.2           \\
 \hspace{3mm}Family, Friends and Others                   & 8           & 36.4           \\
 \hspace{3mm}Friends Only                         & 3           & 13.6           \\
  \hspace{3mm}Friends and Others                   & 4           & 18.3           \\

 \bottomrule
\end{tabular}
\end{table}

\noindent
\textbf{App Tasks:} 
During the field study, our participants were asked to explore different parts of the CO-oPS app through a set of tasks each week. These tasks prompted them to become familiar with CO-oPS features and introduced them to the goal of collaboratively managing mobile privacy and security. Table-\ref{tab:app-tasks} depicts the weekly tasks. For example, Week 1 tasks asked participants to become aware of their own mobile privacy and security decisions, whereas Week 2 tasks asked them to perform oversight of others in their community. Participants could check off completed tasks in the app to remove them from their task list, but we otherwise did not track or require completion to continue in the study. \\

\begin{table}[]
 \footnotesize
 \caption{Weekly App Tasks}
  \label{tab:app-tasks}
\begin{tabular}{p{8cm}}
\hline \hline
\textbf{Week-1:} 1) Review your own apps from the “Discovery” page > “Installed” tab. Hide the apps that you do not want others to see. 2) Review your community’s apps from the Discovery > “All Apps”. Check if you have any uncommon apps that no one else is using. 3) For the apps you have in common with others, compare the permissions you granted but others denied.\\

\textbf{Week-2:} 1) Read the weekly pro tip and add a comment there. 2) From the “People” page, review one of your community member’s apps. Check if there are any apps or permissions that may not be safe. 3) Send a message to warn them about unsafe apps or permissions. \\
  
\textbf{Week-3:} 1) Read the weekly pro tip and add a comment there. 2) Review your own apps and permissions and check if any granted permissions may be unsafe. Consider changing those permissions. 3) Review the apps and permissions of someone in your community. Let them know if they have any unsafe permissions.\\
  
\textbf{Week-4:} 1) Check the messages received from your community. Consider changing the apps and permissions accordingly. 2) Review your community members’ apps and check if any unsafe apps or permissions exist. 3) Write a post on the Community feed to warn others about the unsafe apps or permissions found. 
    \\ \hline \hline
\end{tabular}
\end{table}

\noindent
\textbf{Survey Design: }
Each participant completed two Qualtrics surveys (pre-study and post-study) before and after the field study, which contained four constructs: self-efficacy, community belonging, community trust, and community-collective efficacy. The pre-study survey also collected participants’ demographic information, e.g., age, gender, ethnicity, and education. During the field study, participants also completed a shorter Qualtrics survey each week (weekly survey), containing all constructs of the community oversight model. Links to the weekly surveys were delivered through the CO-oPS app, which redirected participants to the Qualtrics web survey. \\

\noindent
\textbf{Follow-up Interview: }
At the end of the field study, we invited participants to an optional 30-minute one-on-one interview session on Zoom to learn about their experience using the CO-oPS app with their community. Fifty-one participants from 18 communities participated in the follow-up interviews. We started the semi-structured interview by asking about mobile privacy and security practices before participating in the study. Next, we asked about their overall experience of using the CO-oPS app. Participants were also encouraged to express their perceived benefits and concerns about different features of the CO-oPS app. Appendix B presents some sample interview questions we asked during the follow-up interviews. The interview sessions ranged from 40-70 minutes and were audio/video recorded.\\

\noindent
\textbf{Data Collection and Analysis: }
The study produced a rich dataset: 1) quantitative data from survey measures, 2) CO-oPS app usage logs, and 3) qualitative data from follow-up interviews. We first categorized the survey responses as pre-study, week-1, week-2, week-3, week-4, or post-study, depending on the timestamps of the survey completion. Then, we verified the construct validity of our measures using Cronbach's alpha \cite{cronbach_coefficient_1951} and created sum scores to represent each construct. Next, we conducted Shapiro–Wilk tests and found that the sum scores of the constructs were not normally distributed ($ps\!<\!.01$). Therefore, we performed \edit{the non-parametric Wilcoxon rank-sum} test to identify significant differences between the pre-study and post-study measures (Table-\ref{tab:ttest}). \edit{We also present the descriptive statistics for
each pre- and post-study survey item of the newly
developed constructs (Appendix D)} 


We instrumented the CO-oPS app to log participants’ usage data. We also stored the list of the apps installed and permissions granted/denied during the installation of the CO-oPS app and at the end of the field study. We analyzed the usage log to identify how and at what frequencies participants utilized different features of the CO-oPS app. We also analyzed the pre- and post-study app/permissions lists to investigate the changes made to the apps and permissions during the study. Due to some technical issues with the CO-oPS logging feature, we could not log the in-app activities of the first seven communities. Therefore, the app usage data was received from only the last fifteen communities (N = 68 participants). 

We qualitatively coded our interview data using inductive analysis techniques \cite{Fereday_Demonstrating_2006} to understand how participants perceived the CO-oPS features that tie to the constructs we were measuring. Thus, our qualitative analysis complemented the quantitative results from our surveys. We first familiarized ourselves with our data by reading through each transcript and then template-coded our data based on the community oversight concepts described in Section 4. Specifically, we coded for 1) the level of transparency on the information shared by them or others, 2) the types of information they felt helped raise their awareness about the community's mobile privacy and security practices, 3) the level of trust of one another's privacy behaviors and advice, 4) how and whether they would individually participate in such a community, 5) how participants discerned community participation, 6) the trust and belonging they felt with their communities, and 7) their individual and community-level capacity to manage mobile privacy and security. The first author worked closely with three researchers to code the data iteratively and formed a consensus among their codes. The remaining authors helped guide their analyses and interpretation of the results. Appendix C presents the codes and illustrative quotations for each analysis theme. \\


\begin{table*}[t]
 \footnotesize
 \centering
 \caption{CO-oPS App Activities}
  \label{tab:usagelog}
\begin{tabular}{lcccl}
 \hline
\textbf{Activity Types}                   & \textit{\textbf{N = 68}} & \textit{ \textbf{100\%}}  & \textbf{Average*} & \textbf{Apps/Permissions Types}
 \\\hline \hline
 Hide installed apps   & 34           & 49\%           & 6 & Games, video streaming, banking, online shopping \\
Review own app permissions   & 61           & 87\%         & 23  &  Games, online shopping, social media, mobile payments  \\
Review others' app permissions  & 46          & 65\%        & 18 &  Social media, banking, games    \\
Send private messages & 51           & 74\%           & 1 & Apps: games, social media; Permissions: locations, camera, microphones, contacts      \\
\hline

 \multicolumn{5}{l}{* Average number of activities per user (on apps, permissions, or messages)} \\
\end{tabular}
\end{table*}

\section{Results}
On average, participants spent 32 minutes in the CO-oPS app over four weeks, ranging from 19 minutes to 1 hr 31 minutes. Table-\ref{tab:usagelog} summarizes the activity types that participants performed with the CO-oPS app. Table-\ref{tab:ttest} summarizes Cronbach's alpha, means, standard deviations, skewness, and kurtosis of each construct measured during the study. All Cronbach's alphas were greater than 0.80, which suggests good internal consistency of our measures. Next, we tested for within-group differences in the constructs based on whether the participants completed it at the start or end of the study. We will discuss each measure below, along with the corresponding findings from the qualitative data and usage logs related to each construct.\\

\begin{table*}[t]
 \footnotesize
 \centering
  \caption{Descriptive Statistics and Wilcoxon Rank-Sum Tests of Pre and Post-Study Responses to the Constructs}
  \label{tab:ttest}
\begin{tabular}{lccccccccclc}  
\hline
\textbf{Constructs}   &  & \multicolumn{4}{l}{\textbf{Pre-Study}} & \multicolumn{4}{l}{\textbf{Post-Study}} &  &   \\ \cline{3-10} 
 & \textit{$\alpha$} & \textit{M1} & \textit{SD1} & \textit{S1} & \textit{K1} & \textit{M2} & \textit{SD2} & \textit{S2} & \textit{K2} & \textit{V-val} & \textit{p-val} \\ \hline
 \underline{\textbf{Community Oversight Model:}} & & & & & & & & & & & \\

\textbf{Transparency} & 0.88 & 4.07 & 0.80 & -0.74 & 0.70 & 
4.36 & 0.67 & -0.74 & -0.49 & 1054* &  \textbf{0.010} \\

\textbf{Awareness} & 0.82 & 3.95 & 0.79 & -0.76 & 0.84 & 
4.37 & 0.68 & -0.72 & -0.45 & 1110.5*** &  \textbf{<0.001} \\

\textbf{Trust} & 0.90 & 3.61 & 0.80 & -0.29 & 0.11 & 
4.28 & 0.85 & -0.81 & 0.11 & 633.5*** &  \textbf{<0.001} \\

\textbf{Individual Participation} & 0.87 & 3.78 & 0.83 & -0.67 & 0.59 &
4.23 & 0.73 & -0.89 & 0.17 & 897.5*** &  \textbf{<0.001}\\

\textbf{Community Participation} & 0.88 & 3.86 & 0.80 & -0.56 & 0.24 &
4.18 & 0.83 & -0.92 & 0.23 & 1002** &  \textbf{0.002} \\ \hline

\textbf{Community Trust} & 0.87 & 4.03 & 0.87 & -0.85 & 0.45 &
4.22 & 0.71 & -0.78 & -0.03 & 1412* &  \textbf{0.048} \\

\textbf{Community Belonging} & 0.91 & 4.09 & 0.67 & -0.53 & -0.52 & 4.20 & 0.68 & -0.60 & -0.88 & 1899 & 0.209 \\

\textbf{Community Collective Efficacy} & 0.91 & 3.80 & 0.78 & -1.09 & 2.56 & 4.12 & 0.68 & -0.44 & -0.36 & 1287*** & \textbf{<0.001} \\

\textbf{Self Efficacy} & 0.85 & 3.95 & 0.64 & -0.58 & 0.79 & 4.32 & 0.61 & -0.80 & 0.27 & 1021*** & \textbf{<0.001} \\

\bottomrule
*p\textless{}.05; **p\textless{}.01; ***p \textless .001 & & & & & & & & & & & \\

\end{tabular}
\end{table*}

\noindent
\textbf{Transparency: }
As shown in Table \ref{tab:ttest}, participants reported higher ($p\!=\!.010$, \edit{M1=4.07, M2=4.36}) levels of transparency (an individual's perception of whether CO-oPS gave them a transparent view of the apps installed and permissions granted on their community’s mobile devices) at the end of the study. Hence, this result supported our hypothesis (H1). Our qualitative results also confirmed that almost all participants felt that CO-oPS made their community's mobile privacy and security decisions visible to them. Three-quarters of the participants interviewed (76\%, N=39) explicitly said they liked the CO-oPS feature that let them \textbf{check own apps and permissions}. Participants often mentioned that having the ability to see all their installed apps on one screen provided them transparency of their app usage. Two-thirds of the participants (67\%, N=34) also brought up the \textbf{visibility of others' apps and permission}. To this end, they said reviewing others' apps and permissions provided them a sense of purpose for using CO-oPS with their communities. Interestingly, while these participants appreciated the ability to review one another's apps and permissions, they often referred to the importance of the CO-oPS app-hiding feature because it made them feel less intrusive to others' privacy. As such, C18P1 said, \textit{"Because some of the apps can be hidden if someone likes, that gives me the feeling of a relief when I see others' [apps installed].”}. However, some participants (25\%, N=13) believed that this app-hiding feature \textbf{defeats the main purpose} of CO-oPS.

Some participants, on the other hand, perceived transparency as a two-way privacy violation, e.g., the privacy of themselves and the privacy of others. For example, more than one-third of the participants (38\%, N=20) felt that \textbf{their personal privacy was being violated} as others, who were not close, could see their personal information (e.g., installed apps and permissions). Some participants (27\%, N=14) also specifically said \textbf{they might forget to hide an app after installation}, which could leave their apps visible to others. On the contrary, one-fourth of the participants (25\%, N=13) felt that this transparency of others' apps and permissions \textbf{can be privacy-invasive to others} as well. C11P2 said, \textit{“While using the app, my friends and I discussed privacy more than security because we can see the apps on our friend phones and I think that's not a good thing.”}. 



The results from the log analysis (Table-\ref{tab:usagelog}) also supported the above concerns. For instance, around half of the participants (49\%, N=34) hid one or more of their installed apps from their communities. Participants, on average, hid six mobile apps ranging from one to 17. The most frequent types of apps that participants hid were games, video streaming apps, banking apps, and online shopping. \\

\noindent
\textbf{Awareness}: Participants overall reported a higher ($p\!<\!.001$, \edit{M1=3.95, M2=4.37}) level of awareness (individual's perception of whether CO-oPS made them more aware of their community’s mobile privacy and security decisions) after the study. Hence, this result supports our hypothesis (H2). Our qualitative results showed that using the CO-oPS app helped participants raise their overall awareness of mobile privacy and security issues, along with their awareness of one another's privacy and security practices. For example, almost all participants (94\%, N=48) felt that they became more \textbf{aware of mobile privacy and security issues} because CO-oPS enabled them to focus on permissions. They also became more aware of which of their personal information was being accessed by their installed apps. For example, C15P1 said, \textit{“It just makes it more obvious. It's very focused on permissions. So I think having that focus, it's very beneficial. People in the community, I see are now more concerned... for their permissions specifically. I totally see how it changed our perspectives.”}. Some of these participants (39\%, N=20) often brought up the \textbf{weekly pro tips} they got on the CO-oPS app as it helped them increase their awareness regarding mobile privacy and security in general.

Most participants (86\%, N=44) also said they became more aware of whether \textbf{a permission is necessary for an app}. They often mentioned that comparing their own app permissions with others helped them increase this knowledge. Almost all of these participants (82\%, N=42) also mentioned that it helped them \textbf{keep track of their own apps}, as they found more installed apps on CO-oPS' Discovery page than they were aware of. They also often mentioned some granted permissions found on the CO-oPS app (e.g., microphone, camera, location, contacts, etc.) that they did not remember granting. 

Around half of the participants (57\%, N=29) said they \textbf{became more aware of their community members' privacy and security behaviors}. Here, they mostly mentioned one or two people in their community whose apps they could keep an eye on to ensure their safety. Lastly, some participants (39\%, N=20) also said that they appreciated the CO-oPS feature that \textbf{informed everyone about the permission changes} made by any member as this helped them decide whether to imitate that change. Around one-fifth of the participants (18\%, N=9) felt that CO-oPS app \textbf{did not make them aware of the app changes} made in the community - the apps community members installed or uninstalled on their phones. This was not a feature we implemented, and these participants felt like it had been overlooked and desired that awareness. 


The findings from our log analysis (Table-\ref{tab:usagelog}) are reflective of the quantitative results. For example, most of our participants (87\%, N=61) checked their own app permissions during the study. Participants, on average, reviewed the permissions of 23 apps and primarily explored the permissions of apps that are about gaming, online shopping, social media, and financial payments. Alongside reviewing their own apps and permissions, more than two-thirds of the participants (65\%, N=46) reviewed others’ app permissions. On average, they explored 18 app permissions of their community members. The most common types of apps being reviewed were social media, banking, and gaming apps. \\

\noindent
\textbf{Trust}: 
Similar to the above two constructs, the post-study responses saw a higher ($p\!<\!.001$, \edit{M1=3.61, M2=4.28}) level of trust (individual's perception of whether CO-oPS helped them foster trust in one another’s mobile privacy and security decisions) among the community members. This result confirmed our hypothesis (H3). Almost half of the participants (51\%, N=26) said they \textbf{found the advice provided by their community was dependable}. They were overall appreciative of the feedback and guidance they received from more tech-savvy community members, as it helped them learn more about risky apps and unnecessary or dangerous permissions. However, the trust did not always extend to all community members. For example, C18P1 said: \textit{"In this app, you're trusting each other's decisions. But for me, in this community, only [Name] is more tech-savvy. And most of the people are not. And these decisions are not always well-informed, right? So, I follow only [Name] to check what he has."}.


Conversely, participants felt that trusting others' privacy and security practices might be challenging in some cases. Around half of the participants (49\%, N=25) said \textbf{some of their community members were less knowledgeable} about mobile privacy and security issues. Therefore, they could not trust those people's mobile privacy and security decisions to learn from. As C11P3 said, \textit{“I don't think they were much of aware. They do not care of all this, you know, privacy and security stuff,... so I am not sure I followed them, their permissions and stuff.”}. Interestingly, they also often mentioned that those with less knowledge were \textbf{less tech-savvy} (37\%, N=19) in general. \\

\noindent
\textbf{Individual Participation: }
Participants reported a higher ($p\!<\!.001$, \edit{M1=3.78, M2=4.23}) level of individual participation (perception of whether the CO-oPS app helped individuals participate in their own and others' mobile privacy and security decisions) after using the CO-oPS app for four weeks. This supported our hypothesis (H4). Our qualitative results also revealed that participants overall took the initiative to change their apps and permissions and also provided their oversight to others. Notably, 
more than two-thirds of the participants (67\%, N=34) said that they \textbf{made changes to their own apps and permissions}. Participants often said that they made these changes after reviewing their own permissions and identifying the unnecessary or concerning ones by themselves. Some other participants said that comparing their own app permissions with their community's inspired them to change their app permissions. Some of these changes were made because of feedback received from other community members. C02P1 said, \textit{“I did some changes. I denied some of my permissions. [Name] asked me to remove the microphone from one of the apps I use for workouts. I have removed it now. ... also, you can always just check and then you just have to learn what permissions are suspicious and what are necessary.”}.

Next, more than one-third of the participants (41\%, N=21) explicitly said that they \textbf{provided feedback to their community members} to warn about the apps that they thought might be risky or the permissions granted that might be a cause of privacy concerns. To provide feedback, participants did not just use the CO-oPS messaging feature, they also mentioned using other media, e.g., text messages, social media private messages, phone calls, or talking in person.


Log results (Table-\ref{tab:usagelog}) demonstrate that individuals did provide oversight during the study. We found that 74\%, N=51 participants sent messages to someone in their communities, where twelve messages were about warnings regarding risky apps (games, social media). Thirty-five messages contained warnings regarding specific app permissions they found on their community members’ phones. They mostly provided feedback about location, camera, microphones, and contact permissions. For instance, C09P1 messaged C09P3: \textit{“You’re granting Douyin a ton of permissions. maybe we should keep the Chinese spyware to a minimum.”}

However, some participants expressed a number of factors that reduced their motivation to participate. More than one-third of the participants  (41\%, N=21) believed that they \textbf{were less tech-savvy than others in their community} and therefore they doubted their feedback would be useful to others. Interestingly, some participants (39\%, N=20) felt that the \textbf{people who participated with them were not close} and therefore they did not care about those people's mobile privacy and security. A few participants (29\%, N=15) expressed that they \textbf{had very few mobile apps installed} on their devices, and so, they did not need to be concerned about mobile privacy and security. Ironically, some of these participants also believed that they did not have anything to be concerned about because the personal information that is stored in their mobile phones is not very sensitive in nature. A few also felt that their information was already leaked by some online entities and so it was too late to start caring about mobile privacy and security. As such, C14P2 said: \textit{“I don't see the point now because you can't just control what they [apps] already stole from you. I use very few apps, and all my data is already out there}."\\


\noindent
\textbf{Community Participation: }
The community participation measure (individual's perception of whether the CO-oPS app enabled the community to help one another make their mobile privacy and security decisions) increased ($p\!=\!.003$, \edit{M1=3.86, M2=4.18}) over the duration of the field study. This confirmed our hypothesis (H5). More than three-fourths of the participants (78\%, N=40) said the CO-oPS app \textbf{allowed them to learn from their community} regarding mobile privacy and security management and exchange their knowledge regarding app safety and privacy. Most of these participants (71\%, N=36) also mentioned that using CO-oPS helped them \textbf{initiate more open discussions} regarding mobile privacy and security in their community than ever before. They said these discussions most often took place offline when they saw one another in different social gatherings. Around half of the participants (53\%, N=27) specifically discussed \textbf{receiving feedback and advice from their community}. C17P1 said, \textit{“I mean, offline, or virtually, we kind of worked together, we talked, we get each other’s knowledge. But that also happened with the co-ops app, that there were so many options to get in touch with each other by that messaging or, notifying them, or community discussion... I will say it kind of, we helped one another learn as a team.”}. 

However, some participants said the CO-oPS app might not help increase community participation when the members are extremely or not particularly tech-savvy. One-third of the participants (31\%, N=16) envisioned that \textbf{when the community members are less tech savvy}, they might not be able to provide oversight to each other. On the other hand, 27\% of the participants (N=14) said that their \textbf{entire community was very tech-savvy and well aware of the mobile privacy and security issues}, and therefore they did not find it necessary to engage in discussion or exchange feedback with one another.  C11P5 said: \textit{“My community is from a computer science background. I think we are already aware of these things. So, we don't need others' advice.”}. \\

\noindent
\textbf{Community Trust and Belonging: } 
While community trust increased over the course of the study ($p\!=\!.048$, \edit{M1=4.03, M2=4.22}), the difference between community belonging was not statistically significant ($p\!=\!.209$, \edit{M1=4.09, M2=4.20}). Thus, \edit{hypothesis (H6) is supported,} but (H7) is not supported. In our qualitative analysis, we found that all of our participants (100\%, N=51) said they \textbf{personally knew each member} of their communities. Most of our participants (86\%, N=44) mentioned \textbf{having close relationships}, e.g., family members, friends, co-workers, and neighbors, with some members of their communities. Thus, using CO-oPS did not appear to bring groups closer together.

However, perceptions of trust and community relationships were still important in how individuals interacted with each other in CO-oPS. Around half of the participants (47\%, N=24) said that they \textbf{had trust in their community that their apps and permission information would not be misused}. One-fourth of our participants (24\%, N=12) said they had peace of mind because they would \textbf{rely on their community members} who would actively monitor their mobile privacy decisions and warn them if anything is found concerning. Here, we often noticed that participants referred to some specific community members, not the entire community, who they would rely on. C02P1 said, \textit{"With [Name] in my group, at least I know that if he saw something he didn’t think wasn’t proper, he will definitely let me and my husband know...We have that kind of relationship, so we know we can trust him.”}.

However, a few participants felt that sharing the apps installed might cause some security issues due to the lack of trust in certain community members. For example, a few participants (18\%, N=9) envisioned \textbf{security concerns in sharing their financial apps}, such as banking or mobile payments, with their community. They often brought up hypothetical scenarios of a family member (e.g., children) knowing what apps they have installed, who would somehow get access to their phone, log in to their financial apps, and transfer money. A couple of participants also imagined situations when community members \textbf{might judge or bully them because of their choice of gaming apps}.\\


 
\noindent
\textbf{Self-efficacy: }
Our participants reported higher levels of self-efficacy (individual's capacity to manage their mobile privacy and security) at the end of the study ($p\!<\!.001$, \edit{M1=3.95, M2=4.32}). This confirmed our hypothesis (H8). Most participants (80\%, N=41) said they \textbf{gained confidence in managing their mobile privacy and security}, particularly by reviewing their installed apps and granted permissions and identifying whether there is anything concerning. C10P1 said, \textit{"So, I can now think through it, like what is the purpose of this permission? Like if the permission conflicts with the purpose of the application, I can just turn it off. You see, this is new. I now can differentiate what's necessary or what's not."} Interestingly, more than half of the participants (57\%, N=29) said they now have \textbf{become more knowledgeable about changing permissions}, mostly because they could easily navigate to the app permission settings from the CO-oPS apps. This perception was not universal, though. Around one-third of the participants (31\%, N=16) also said they \textbf{already had the ability to manage their own apps and permissions} prior to participating in this study, and they never reached out to others for help.\\

\noindent
\textbf{Commmunity Collective Efficacy: }
Participants reported higher community collective efficacy (individuals' belief that their community can co-manage mobile privacy and security) at the end of the field study ($p\!<\!.01$, \edit{M1=3.80, M2=4.12}). This confirmed our hypothesis (H9). Reflecting this, most participants (88\%, N=45) felt they could easily \textbf{reach out to their community and work together as a team} for their mobile privacy and security decisions. Most of these participants (67\%, N=34) mentioned that they \textbf{have at least one person in the community they could reach out to ask questions} about whether an app was safe to use or a permission should be allowed. C03P5 said, \textit{“When I'm giving permissions, I now can tell that could be the things that are needed for a discussion. I do go to [Name] to ask what he thinks. what he thinks the permission is needed or not needed for the app. I do my permissions like this now.”}\\

\noindent
\textbf{Behavioral Impact:}\\
Our log analysis results provide further insights into participants' overall behavioral changes regarding mobile privacy and security. We found that 87\% of the participants (N=61) changed at least one of their app permissions during the study. Participants, on average, changed 29 permissions, where all permissions were changed to “deny.” They mostly turned off the permissions accessing their location (approximate and precise), camera, storage, and contacts. For instance, C15P4 changed the Location (Approximate) permissions of Chase, Snapchat, and Gyve apps installed on his phone. However, participants did not show a similar decrease in the number of apps they had on their phones. Around 78\%, N=53 participants installed new apps, whereas only a few participants (16\%, N=11) uninstalled any apps. Participants, on average, installed two new apps, where the most common types of apps were mobile payment, banking, online shopping, social media, and games. On the other hand, the participants who uninstalled their apps mostly discarded gaming apps along with a few spiritual, fitness, and dictionary apps from their phones. Perhaps learning what apps others in their communities were using provided participants with ideas for additional apps they would be interested in.\\

\section{Discussion}
\textcolor{black}{While our prior work conceptually proposed community oversight as a mechanism for supporting privacy and security management \cite{chouhan_co-designing_2019}, }this work is the first field study \edit{to empirically examine the real-world feasibility of implementing community oversight as a mechanism for co-managing} mobile privacy and security \edit{among trusted groups}. Our results largely confirm what was envisioned in that prior work: that community oversight does have the potential to help people help each other when it comes to decisions about mobile apps and app permissions \cite{chouhan_co-designing_2019}. Users' perceptions of their own and their community's capabilities to manage their mobile app privacy and security increased as a result of the study. The majority of participants modified their permissions, reducing what they were sharing with apps, and stated that their awareness of permissions and mobile apps also increased. Below we further discuss our overarching findings and their implications for the design of community oversight mechanisms.\\


\noindent
\textbf{Building Community Collective Efficacy } \\
The goal of the CO-oPS app, as with many collaborative systems, is to build and support the collective capacity of groups to work together to achieve a common goal, in this case, to manage apps and app permissions. Thus, building community collective efficacy for mobile privacy and security is the primary end goal of CO-oPS. To that end, we believe our study was successful. The interview comments suggest that the community oversight mechanism helped our participants increase their ability to support each other in their mobile privacy and security decisions. Participants mentioned their change to a more collaborative perspective: the app facilitated knowledge sharing amongst their community and an ability to rely on others to help in decision-making \cite{badillo2020towards}. 
\textcolor{black}{Our results also provide an empirical validation of the components of the community oversight model \cite{chouhan_co-designing_2019}.} Again, both survey and interview results demonstrated the roles of transparency, awareness, trust, and participation in providing community oversight. Future work could examine what factors are most related to community collective efficacy and thus are most important to provide in a community oversight mechanism. \\


\noindent
\textbf{Role of Tech Expertise} \\
One of the key themes was that the level of tech expertise among community members plays a key role in bolstering or hindering community oversight. For instance, our participants expressed concerns about the potential lack of participation in communities when most members are sufficiently tech-savvy or knowledgeable about mobile privacy and security. Others expressed concerns about there being a lack of knowledge in their communities and less trust in the decisions of those with less expertise. Kropczynski et al. \cite{kropczynski_towards_2021} also noted the importance of those with tech expertise in older adult communities for spreading privacy and security knowledge, even among those with low self-efficacy. This suggests that community oversight mechanisms may be most beneficial and appropriate when there are asymmetrical relationships among the community members in such a way that some community members need support while other members could provide that support. A key challenge is then how to incentivize those with sufficient expertise to participate in such communities, particularly to help community members they are not as close to or not already providing tech care to \cite{kropczynski_examining_2021}. 

\edit{However, when this asymmetry in expertise combines with a power imbalance, which is often seen in families, the collaborative joint oversight might cause tension. Akter et al. \cite{akter_from_2022} demonstrated that although teens had more expertise than their parents, they did not feel empowered to oversee their parents because of the existing power hierarchies. In families, parents often use parental control apps, a more restrictive approach that fosters monitoring and surveillance to ensure teens' mobile online safety, privacy, and security. Teens often perceive this unidirectional oversight mechanism as overly restrictive and privacy-invasive \cite{wisniewski_parental_2017, ghosh_safety_2018}. Therefore, adolescent online safety researchers emphasize adopting a softer version than parental control or community oversight - a middle ground that allows parental oversight with bidirectional communication and teens' self-regulation \cite{wisniewski_parental_2017}. So, the community oversight mechanism might need to incorporate additional features to help such unique types of communities with asymmetries in expertise and power.} \\

\noindent
\textbf{Tensions around Transparency and Privacy}\\
Another common concern was privacy issues arising from transparently sharing apps and permissions with others. While many appreciated such transparency, participants regularly chose to hide certain apps from other people. Some participants found this transparency too invasive and anticipated potential problems resulting from others knowing about what apps they use. Other concerns also arose from being able to determine if the advice given to another was taken or not, based on whether someone's permissions remained the same or changed. These concerns will likely be elevated as community size grows, where communities contain more members who are not close to one another. \edit{A recent study that explored collaborative mobile privacy management among families also found similar results where participants expressed concerns in including extended families with distant relationships \cite{akter_it_2023}. To resolve these tensions,} as with many collaborative systems, users may want more granular controls on who can see what apps and permissions rather than sharing equally throughout the community. \\

\noindent
\textbf{Incentives to Participate}\\
\textcolor{black}{Prior work identified that users might not be motivated to provide oversight to those not close to them \cite{chouhan_co-designing_2019}} or those outside of existing care relationships such as between parents and teens \cite{akter_from_2022}. Indeed, some of our participants expressed similar sentiments and were not concerned about the decisions made by those not close to them. Despite this, the majority of participants did perform oversight, and many interviewees described discussions and behaviors that were sparked as a result of that oversight. Yet with some incentives, participating in a user study, in this case, individuals performed the oversight, benefiting other community members. Thus a key question remains as to how to incentivize such oversight to different community members and how those incentives may need to change over time.\\

\noindent
\textbf{Implications for Design}\\
Our results demonstrate how features that provide transparency and awareness and support trust between community members are essential components of community oversight. Mechanisms must also enable and encourage individual and community participation in the collaborative efforts of privacy and security management. Our results provide further insights into the features and mechanisms needed in a tool for communities to participate in collaborative oversight of their mobile privacy and security.

\textit{Making Privacy Features Visible:} While the CO-oPS app had a feature that allowed users to hide any of their installed apps from others, it often failed to provide users with a sufficient sense of privacy. This may be because they were not well aware of this feature or were unsure how well it functioned. Participants also reported concern over forgetting to hide apps as they install new ones. Thus, mechanisms to keep users aware of this app-hiding feature will be necessary. \edit{Das et al.  \cite{das_the_2015} and DiGioia and Dorish \cite{dourish_social_2005} also emphasized the importance of visibility so that users can be aware of the availability of the security feature and adopt it.} To help users be aware of this feature, users can be prompted regularly or upon installing new apps to ask if they would like to hide. If community members hide too many apps, however, oversight will be more limited. Thus, designers should also explore additional privacy features that can protect an individual's privacy while still allowing useful sharing to the community. 

\textit{Raising Mobile Privacy Knowledge:}
One of our findings suggested that participants would not trust the mobile privacy and security behaviors of people who were less knowledgeable. \edit{This suggests that collaborative decision-making would not effectively function when there is little trust within the group. Increasing trust within communities may be very challenging, and how to do so remains an important open question.} 
\textcolor{black}{In \cite{chouhan_co-designing_2019}, participants also envisioned such situations and recommended including external expert users whom the community members can turn to for guidance when they do not have the necessary expertise.} \edit{Several other networked privacy researchers also demonstrated the need for knowledgeable expert stewards \cite{bonneau_alice_2009, murthy_individually_2021, bonneau_privacy_2009}.} Therefore, we recommend app designers explore ways to include mobile privacy and security experts in communities. Another possibility is, rather than bringing experts into the community, to raise the expertise of certain motivated community members. This could include nudges towards additional information or resources, possibly personalized to those most amenable to such additional knowledge. 

\textit{Increasing Community Participation:}
We found that our participants expressed several concerns about community motivations to provide oversight to one another. Individuals and communities, as a whole, need incentives to utilize a community oversight mechanism and continue to support each other \cite{aljallad_designing_2019, poole_computer_2009} in their knowledge-sharing and decision-making. \edit{Such needs for incentivizing individual participation in communities to support collective participation were also suggested by Watson et al. in \cite{watson_we_2020} and Moju-Igbene et al. in \cite{moju_Igbene_how_2022}).} Therefore, community oversight mechanisms need to include features that encourage such engagement and make the engagement of others apparent. For example, community members can be notified of any new apps installed or permissions granted on anyone's phone. Moreover, nudges could remind community members to review random members' apps and permissions. Additionally, lightweight feedback features might also help users to engage more. For instance, instead of messaging, users might prefer just to flag unsafe apps/permissions to notify others quickly. \\

\noindent
\textbf{Limitations and Future Work}\\
We would like to highlight the limitations of our study that should be addressed in future work. First, our sample was skewed toward Asian adults, most of whom completed college and graduate-level education. Therefore, our results may not be generalizeable to other communities of different ethnicity, education, and age groups. Future work should explore communities with broader demographics, ethnicity, and socio-economic status \cite{madden_privacy_2017}. Another limitation is that we asked our initial participants to form their communities with people they knew, which sometimes led to groups where everyone did not have strong bonds with each other. This may have led them to evaluate our app differently than if we studied with communities of families or close friends only. However, this also provided important insights into the importance of community trust in fostering oversight. \textcolor{black}{Future work should examine how factors of group structure and relationships, including group size and varying levels of expertise, impact the motivation of participation and oversight activities of community members.}

\edit{Although our qualitative results suggested that} the CO-oPS app supported all necessary components of community oversight, this does not imply that our participants perceived usefulness, ease of use, and behavioral intent to adopt \cite{davis_perceived_1989}. \edit{This is because they used the app, as we requested, to perform various tasks as part of the study.} Therefore, in future studies, we would want to evaluate its usability to address users' experience issues and measure technology acceptance \cite{davis_perceived_1989} to identify how to design for widescale adoption of an app to help people collaborate with their loved ones to manage mobile privacy and security. \textcolor{black}{Lastly, the study design did not include a control condition, which means that any effects from the community oversight mechanism cannot be differentiated from changes that may have occurred through using the app, such as increased attention on app permissions and privacy and security.} \edit{Therefore, the results cannot conclusively demonstrate a causal relationship between the usage of CO-oPS with communities and the dependent variables we analyzed. However, our qualitative insights provide evidence that some of the positive effects could be attributed to using the CO-oPS app. Moreover, there might be a survivorship bias effect in our results, as those who dropped out did not perceive any benefits to the app. Future research should investigate whether the same findings would hold for control groups 
and prevent potential survivorship bias. }


\section{Conclusion}
Managing mobile privacy and security as an individual is hard. We believe community oversight is one potential social mechanism that can allow community members to exchange help regarding their mobile privacy and security decisions. Our CO-oPS app was developed to evaluate this idea of community oversight in building community collective efficacy for groups managing their mobile privacy and security together. Our results provide empirical evidence that community oversight can potentially have an impact on individuals and communities alike. Given the continued proliferation and adoption of smartphones and mobile apps, we believe apps that facilitate community oversight are an essential tool for communities to help one another keep their personal information safe and secure. We will continue to build upon this work to examine how we can help people successfully co-manage mobile privacy and security within their communities.


\section*{Acknowledgments}
We acknowledge the contributions of Nikko Osaka, Anoosh Hari, and Ricardo Mangandi, in the CO-oPS app development. We would also like to thank the individuals who participated in our study. This research was supported by the U.S. National Science Foundation under grants CNS-1814068, CNS-1814110, and CNS-2326901. Any opinion, findings, and conclusions or recommendations expressed in this material are those of the authors and do not necessarily reflect the views of the U.S. National Science Foundation.


\bibliographystyle{plain}
\bibliography{usenix2022_SOUPS}
\newpage
 \onecolumn
\begin{appendices}
 \appendix
\section{Survey Scales}

\noindent
\textbf{Community Oversight Model Constructs:} \edit{(Derived from Chouhan et al.'s conceptual model of Community Oversight \cite{chouhan_co-designing_2019})}\\
 \textbf{Transparency}\\
1. The app gave me a transparent view into the apps installed and permissions granted on my own mobile device.\\
2. The app gave me a transparent view of the apps installed and permissions granted on the mobile devices of others.\\
3. The app gave us all a transparent view of the apps installed and permissions granted on the mobile devices of our community.\\
\textbf{Awareness}\\
1. The app made me more aware of my own mobile privacy and security decisions.\\
2. The app made me more aware of the mobile privacy and security decisions of others.\\
3. The app increased overall awareness of the mobile privacy and security decisions of our community as a whole.\\
\textbf{Trust}\\
1. The app helped me foster trust in the mobile privacy and security decisions of others in my community.\\
2. The app helped others in my community foster trust in my mobile privacy and security decisions.\\
3. The app helped foster trust in the mobile privacy and security decisions of our community as a whole.\\
\textbf{Individual Participation}\\
1. The app helped me make privacy and security decisions for myself.\\
2. The app helped me be involved in others' privacy and security decisions.\\
3. The app helped individuals in the community participate in privacy and security decisions of our community.\\
\textbf{Community Participation}\\
1. The app enabled me to participate in a community that helps one another regarding our mobile privacy and security decisions.\\
2. The app enabled others to participate in a community that helps one another regarding our mobile privacy and security decisions.\\
3. The app enabled the community to help one another regarding our mobile privacy and security decisions.\\

\noindent
\textbf{Community Trust} \edit{(Derived from Chouhan et al.'s conceptual model of Community Oversight \cite{chouhan_co-designing_2019})}\\
1. I trust others in my community to protect my private information.\\
2. I trust others in my community to give me advice about mobile privacy and security.\\
3. Others in my community trust me to protect their private information.\\
4. Others in my community trust me to give them advice about mobile privacy and security.\\

\noindent
\textbf{Community Belonging} \edit{(Pre-validated by Carroll et al. \cite{carroll_community_2003} and Sarason et al. \cite{sarason1974psychological})}\\
1. I can get what I need in this community.\\
2. This community helps me fulfill my needs.\\
3. I feel like a member of this community.\\
4. I belong in this community.\\
5. I have a say about what goes on in this community.\\
6. People in this community are good at influencing each another.\\
7. I feel connected to this community.\\
8. I have a good bond with others in this community.\\

\noindent
\textbf{Self-Efficacy} \edit{(Pre-validated by Kropzynski et al. \cite{kropczynski_towards_2021} based on a modified version from Bandura \cite{bandura1982self})}\\
1. I know that if I worked hard to learn about mobile privacy and security, I could make good decisions. \\
2. Mobile privacy and security decision-making is not too complicated for me to understand. \\
3. I think I am the kind of person who would learn to use best practices for good mobile privacy and security decision-making. \\
4. I think I am capable of learning to help others make good mobile privacy and security decisions. \\
5. Given a little time and training, I know I could learn about best practices for good mobile privacy and security decision-making for myself and my community. \\

\noindent
\textbf{Community-Collective Efficacy} \edit{(Pre-validated by Kropzynski et al. \cite{kropczynski_towards_2021} based on a modified version from Carroll et al. \cite{carroll_collective_2005})}\\
1. Our community can cooperate to improve the quality of our decisions about mobile privacy and security.\\
2. Despite other obligations, we can find time to discuss our decisions about mobile privacy and security.\\
3. As a community, we can handle the mistakes and setbacks resulting from our decisions about mobile privacy and security without getting discouraged.\\
4. I am confident that we can be united in the decisions we make about mobile privacy and security that we present to outsiders.\\ 
5. As a community, we provide care and help for one another regarding our mobile privacy and security decisions.\\
6. Our community can leverage outside resources and services for our members to ensure the quality of mobile privacy and security decisions.\\
7. Our community can provide information for people with different interests and needs when it comes to mobile privacy and security decision-making.


\section{Sample Questions of Followup Interview}

  \textit{
    \begin{itemize}[noitemsep,nolistsep]
\item Prior to participating in this study, how did you decide which apps are safe or unsafe to install on your mobile devices? 
\item How did you decide whether to accept or deny a permission request for an app?
 \item Did you ever review the permission lists of the apps installed on your phone? Why or why not? How?
 \item How frequently did people in your community discuss mobile privacy and security issues with one another? 
\item During the study, how frequently did your community members discuss mobile privacy and security decisions with one another? 
\item During the study, how did you communicate with others who were part of your community? 
\item During the study, how did you manage your mobile privacy and security decisions? Did you see any changes compared to prior to the study? Why or why not?
\item Can you explain how and why the app did or did not help provide transparency into the mobile privacy and security decisions of other people in your community? 
\item How and why did the app or did not help raise awareness in your community about mobile privacy and security?
\item How and why did the app or did not enable you and individuals in your community to provide feedback and guidance about others’ mobile privacy and security?
\item How and why did the app or did not help you work together as a community about mobile privacy and security? 
\item Were there any problems or concerns you or others in your community encountered when using the app? 
\item If given the option, would you want to continue using the CO-oPS app after this study? Why or why not? 
\item Who do you think would be benefited the least from using the app and why? Who would be most benefited and why? 
\item Is there anything else that you would have liked the app to do? Any changes you would have liked on how the app currently works?
\end{itemize}
} 

\section{Codebook}

\setcounter{table}{0} 
\renewcommand{\thetable}{A.\arabic{table}}

\begin{center}
 \footnotesize
\begin{longtable}{ |p{3.5cm}|p{12.5cm}|  }

\caption{Codebook} \label{tab:codebook-RQ3} \\

\hline {\textbf{Codes}} & {\textbf{Illustrative Quotations}} \\ \hline 
\endfirsthead

\multicolumn{2}{c}%
{{\tablename\ \thetable{} -- continued from previous page}} \\
\hline {\textbf{Codes}} & {\textbf{Illustrative Quotations}} \\ \hline 
\endhead

\hline \multicolumn{2}{|r|}{{Continued on next page}} \\ \hline
\endfoot
\hline
\endlastfoot

 \multicolumn{2}{c}{} \\[-10pt] 

 \multicolumn{2}{|l|}{\textbf{Transparency}}  \\ \cline{1-2}

 Visibility to own S\&P (security and privacy) \newline (76\%, N=39) & \textit{“So actually using co-ops, like, for me, I got to see the list, like, what the apps, what the actual permission all the apps are using and like, what access they have. Like the list of all at the same place. For me, it was like, good to have this.”} -C11P3   \\ \cline{1-2}
   
 Visibility to others S\&P  \newline (67\%, N=34)
 & \textit{“I think having this app actually made me more, see these things of others, because it made it easier now to check, not only yours, but also other people's security settings.”} -C18P1   \\ \hhline{|==|}

  Violation of own privacy when other's view (38\%, N=20) & \textit{“As some of the community are not someone who not much close, I wasnt that much confident when it came to share my apps and show my things, you know.”} -C08P2   \\ \cline{1-2}
   
 Violation of own privacy when forget to hide (27\%, N=14)
 & \textit{“I didn't want to show a few apps, to my community members, but, as CO-oPS crashed the first time and I had to reinstall it. Then, I forgot to hide those apps. And so I think that is a privacy issue, which most people won't like it.”} -C11P1 \\ \cline{1-2}

   Violation of others’ privacy \newline  (25\%, N=13) & \textit{“While using the app, my friends and I discussed privacy more than security because we can see the apps on our friend phones and I think that's very not a good thing. I did not feel good.”}-C11P2   \\ \cline{1-2}
   
 Defeats the purpose \newline  (25\%, N=13)
 & \textit{“But sometimes, so while people are using some apps and keeping it private to them, they dont share with anyone but yeah, then I think this app wont help much for anyone”} -C06P2   \\ \hhline{|==|}
  

 


 \multicolumn{2}{c}{} \\[-10pt] 
  \multicolumn{2}{|l|}{\textbf{Awareness}}  \\ \cline{1-2} 

 Overall S\&P awareness \newline(94\%, N=48) & \textit{“Sometimes we allow some permissions without understanding what's been packed. So after exploring that CO-oPS, I usually get to think twice about my apps, which really cool, I am more concerned about whether to allow or not allow any permissions to secure your phones. I would say it's very helpful to change my mind. And it helped me to be more careful about my mobile security.”} -C15P1    \\ \cline{1-2}
 
Compare own S\&P with others \newline(86\%, N=44)  & \textit{“I think this is a great feature. Because with this, you are able to see and compare like, if what you are using and what others are using, it is like comparable Or you can just know what you are doing others are not. I guess you can help yourself.”} -C17P1  \\ \cline{1-2}
 
Keep track of own S\&P \newline(82\%, N=42)
 & \textit{“Earlier I couldn't know about what is there and what is not because I thought I had few apps the apps I did install. Then here [on CO-oPS app] I see I have more apps that I did not see it before... I think it helps, it feels like gives you to see what do you have on the phone, and the stuff that are accessed by the apps.”} -C08P1  \\ \cline{1-2}

Aware of others' S\&P \newline(57\%, N=29) & \textit{“So like seeing the option of like, every single app, and then seeing like what's granted and denied, that definitely helped a lot to see what each member, what apps did they have, and also what like permissions they grant. So it helps me realize what they're granting or not granting, so that I need to I help them or not.”} -C02P4  \\ \cline{1-2}

Aware of community's S\&P Changes (39\%, N=20) & \textit{“One of the benefits of it is, on the community section, I can go through my friend's app changes, which permissions of which apps you changed. And I can go ahead and do that and change it and have fun. Okay.”} -C11P1  \\ \cline{1-2} 

Increased awareness from pro tips (39\%, N=20) & \textit{“So on this pro tip section in where you can know the basic information, like basic knowledge that you can just learn from and become careful about the app settings... I think this section talked some senses in us.”} -C06P2 \\ \hhline{|==|}

Doesnt inform community about Apps Changes (18\%, N=9) & \textit{“So, you see in the community, we get to know about the changes for the permissions, but we do not get any community posts for the app installing or installing. I think this is also important. When someone gets rid of an app, everuyone should know, right.”} -C15P1  \\ \hhline{|==|}

 \multicolumn{2}{c}{} \\[-10pt] 
  \multicolumn{2}{|l|}{\textbf{Trust}}  \\ \cline{1-2}

 Trust others’ advice \newline(51\%, N=26)
 & \textit{“[Name] let me reconsider what I am doing, because when he tells me warns me, you are more likely to take it seriously. It'll come to light in your mind for sure. Yeah, I did change some of the things, yeah I think he was right. I see the stuff he warns me about are all good.”} -C11P2    \\ \cline{1-2}
 

Less aware community members (49\%, N=25) & \textit{“I dont think they were much of aware. They do not care of all this, you know, privacy and security stuff, so I am not sure they used it much.”}-C11P3 \\ \cline{1-2}
   
 Less tech-savvy community members  (37\%, N=19) & \textit{"For example, my mom... whenever she goes to the Facebook or YouTube, she asks questions. So, she cant be able to understand these privacy and security, its just so beyond her capacity. So I doubt she would be someone to rely on."}-C16P1   \\ \hhline{|==|}

  \multicolumn{2}{c}{} \\[-10pt] 
 \multicolumn{2}{c}{} \\[-10pt] 
  \multicolumn{2}{|l|}{\textbf{Individual Participation}}  \\ \cline{1-2}

 
Made own S\&P changes \newline(67\%, N=34)  & \textit{“I got rid of some of my permissions. I haven't really thought of that before. Right now it has come to my knowledge that yes, it is a big problem and even scary. But I have that control, if you know what permissions are problematic, and what are necessary, you can always try clean up. Now cleaning up my phone has become a bit of a priority to me.”} -C02P1   \\ \cline{1-2}
   
Provided feedback \newline(41\%, N=21) & \textit{“I  reviewed X’s mobile privacy, I saw he was giving a permission, don't remember which one, then I told him that, allowing that permission is not good. And then I gave some good reasons why this is important to change this or not.}
-C17P1  \\ \hhline{|==|}

 Less tech-savvy  \newline(41\%, N=21)  & \textit{"I dont think anyone needed my advice. I know they are careful, much careful than I am because they all are very savvy.”} -C18P2    \\ \cline{1-2}
 
Others are not close \newline(39\%, N=20) & \textit{I dont think I did much… I would be interested to help someone when I care them, maybe my parents mostly.”}-C14P2   \\ \cline{1-2}
   
Fewer app users \newline (29\%, N=15) & \textit{“I did not use it much. I'm a very minimalist in my apps. So at this point, the apps that I have, I know what I have. My advice to others is use minimal apps and make your life easy..”}-C03P5  \\ \hhline{|==|}
 

  \multicolumn{2}{c}{} \\[-10pt] 
 \multicolumn{2}{c}{} \\[-10pt] 
  \multicolumn{2}{|l|}{\textbf{Community Participation}}  \\ \cline{1-2}

Learned from community \newline(78\%, N=40)
 & \textit{“We could review each other's permissions and we could Share, so we could be careful about our privacy and things. And having your community’s apps and permission in CO-oPS, you can just learn by yourself  like maybe you don't really have to grant this permission.”} -C18P2   \\ \cline{1-2}

Increased discussion in community (71\%, N=36) & \textit{“We had frequent discussions when we had discussions about what kind of security and permissions we have or on each other's phone, or in general the security issues out there. And I think the other day when we met, we were giving away some information. I think we also mentioned some of our apps  are taking unnecessary data. For those apps purpose, the permissions were not necessary. So we asked to turn it off. And I don't know if they did change that, but I did. But yeah, that kind of interaction truly happened among us. And we had we shared opinion and try to suggest each other that this is not right.”} -C06P4   \\ \cline{1-2}

Received feedback from community (53\%, N=27)
 & \textit{“One of the action items we had a task like look through permissions and tell them like, hey, like, maybe you shouldn't do it. I think I received a message from X like, Hey, you have Bose like music app has access to your GPS location for some reason. Oh, wow. Which I did not notice it before. This was like, I really thanked him.”} -C09P2  \\ \hhline{|==|}

Less tech-savvy community (31\%, N=16) & \textit{"I think when your community is not tech savvy, they wont feel the importance of this security and privacy. I can see to be an effective community at least some people must be tech savvy so that they can educate everyone else."} -C16P1   \\ \cline{1-2}

Tech savvy community \newline (27\%, N=14)
 & \textit{“We didnt find it useful, not really, because my community is from computer science background. I think we are already aware of these things. So, we dont need others advice.”} -C11P5 \\ \hhline{|==|}

\multicolumn{2}{c}{} \\[-10pt] 
 \multicolumn{2}{c}{} \\[-10pt] 
  \multicolumn{2}{|l|}{\textbf{Community Trust and Belonging}}  \\ \cline{1-2}
Good relationship with community  (86\%, N=44)
 & \textit{“We live in a same community, so we have a very good relation with the other people like X and all the other four members because we almost live in very close to and very similar minded community. So and I have personally good relationship with X that also drives me to participate in this research. So yeah, we try to go outing and explore things together.”} -C15P1  \\ \cline{1-2}

 Trusted others to keep S\&P Info Private (47\%, N=24) & \textit{“I guess, like the thought that they are my close circles. Like I know sharing my apps with them is safe.”} -C15P1  
  \\ \cline{1-2}

  Depend on the community for S\&P (24\%, N=12) & \textit{“With X in my group at least I know that if he saw something he didn't thought wasn't proper, he will definitely let us know, let me and my husband know.. We have that kind of relation, he, Yeah, he would let us know and he would tell us this just to delete that, we have that kind of relationship so, we know we can trust him, We know that,”} -C02P1 \\ \hhline{|==|}

  Had security concern for sharing S\&P (18\%, N=9) & \textit{“I have my Chase app, if someone on the family, like my sons, know I have this app and can somehow get my phone,... if the app is logged in already, they can just transfer the money immediately.”}-C02P3  \\ \hhline{|==|}

\multicolumn{2}{c}{} \\[-10pt] 
 \multicolumn{2}{c}{} \\[-10pt] 
  \multicolumn{2}{|l|}{\textbf{Self Efficacy}}  \\ \cline{1-2}

Gained confidence in S\&P \newline(80\%, N=41)
 & \textit{“Okay, so, I will say that what is the purpose of this app? Like if it is like Facebook or WhatsApp, then it will use my contacts, my contact information can use or my photos they can use. But why they should go to my phone call manage permission or there will track my other applications permission. That doesn't make sense. So it conflicts with the purpose of this application. See, this is new. I can now differentiate whats necessary or what not."} -C10P1  \\ \cline{1-2}
 
Now know how to change permissions (57\%, N=29)
 & \textit{“So I actually now can use the settings to go directly change the permissions. Its much easier now. It has become like I randomly go check some apps and do changes instantly if I feel like.”} -C12P2 \\ \hhline{|==|}

Already confident in S\&P (31\%, N=16)
 & \textit{“I would say that'd be me. I'm pretty knowledgeable regarding, you know, the whole privacy and phones, I try to be secure about my own apps. Yeah, I think I am very careful with permissions and such. I know how to change things.”} -C22P1  \\ \hhline{|==|}



  \multicolumn{2}{c}{} \\[-10pt] 
 \multicolumn{2}{c}{} \\[-10pt] 
  \multicolumn{2}{|l|}{\textbf{Community Collective Efficacy}}  \\ \cline{1-2}

Felt teamwork for S\&P \newline(88\%, N=45)
 & \textit{“I mean, offline, or virtually, we kind of worked together, we talked, we get each other’s knowledge. We could easily just start a discussion about any apps and permissions stuff... I will say it kind of, we work together in this.”} -C17P1  \\ \cline{1-2}
 
Reached out to community \newline(67\%, N=34)
 & \textit{“ think one thing is that I’m a little more confident of
it now. So, when I’m giving permissions, I now can tell that
could be the things that needed for a discussion. I do go to
[Name] to ask what he thinks would do. what he thinks if
the permission is needed or not needed for the app. I do my
permissions like this now.”} -C03P5 \\ \cline{1-2}



\end{longtable}
\end{center}
\section{\edit{Descriptive Statistics of Community Oversight Construct Items}}
\begin{table*}[tbh]
 \footnotesize
 \centering
  \caption{Wilcoxon Rank-Sum Tests of Pre and Post-Study Responses to the Community Oversight Construct Items}
  \label{tab:ttest_item}
\begin{tabular}{lcccccccclc}  
\hline
\textbf{Constructs}     & \multicolumn{4}{l}{\textbf{Pre-Study}} & \multicolumn{4}{l}{\textbf{Post-Study}} &  &   \\ \cline{2-9} 
 & \textit{M1} & \textit{SD1} & \textit{S1} & \textit{K1} & \textit{M2} & \textit{SD2} & \textit{S2} & \textit{K2} & \textit{V-val} & \textit{p-val} \\ \hline

 \underline{\textbf{Transparency:}}  & & & & & & & & & & \\ 
The app gave me transparency of my apps and permissions  & 4.18 & 0.86 & -0.89 & 0.73 &
4.40 & 0.72 & -0.79 & -0.68 & 716 &  0.08 \\

The app gave me transparency in others' apps and permissions  & 4.02 & 0.90 & -0.85 & 0.97 & 
4.32 & 0.77 & -0.91 & 0.29 & 696* &  \textbf{0.026} \\

The app gave us transparency in whole community's app  & 3.98 & 0.99 & -0.83 & 0.30 & 
4.30 & 0.76 & -0.86 & -0.27 & 654.5* &  \textbf{0.012} \\

 \underline{\textbf{Awareness:}}  & & & & & & & & & & \\ 
The app made me aware of my S\&P decisions  & 4.12 & 0.91 & -1.31 & 2.53 &
4.49 & 0.67 & -1.15 & 1.10 & 462.5** &  \textbf{0.001} \\

The app made me aware of others' S\&P decisions  & 3.81 & 0.99 & -0.74 & 0.42 & 
4.28 & 0.85 & -1.01 & 0.31 & 804** &  \textbf{0.001} \\

The app made me aware of whole community's S\&P decisions   & 3.91 & 0.94 & -0.69 & 0.09 & 
4.23 & 0.84 & -1.01 & 0.52 & 880.5*** &  \textbf{<0.001} \\

 \underline{\textbf{Trust:}}  & & & & & & & & & & \\ 
The app helped me foster trust in others' S\&P decisions  & 3.65 & 0.98 & -0.37 & -0.21 &
4.19 & 0.81 & -0.74 & -0.00 & 405*** &  \textbf{<0.001} \\

The app helped others foster trust in my S\&P decisions  & 3.57 & 0.91 & -0.16 & -0.32 & 
4.02 & 1.02 & -0.96 & 0.39 & 546.5*** &  \textbf{<0.001} \\

The app helped foster trust in whole community's S\&P decisions & 3.60 & 0.84 & -0.43 & 0.17 & 
4.12 & 0.94 & -1.09 & 1.15 & 467*** &  \textbf{<0.001} \\

 \underline{\textbf{Individual Participation:}}  & & & & & & & & & & \\ 
The app helped me make S\&P decisions for myself  & 3.96 & 0.98 & -0.82 & 0.39 &
4.37 & 0.79 & -1.43 & 2.84 & 727.5** &  \textbf{0.002} \\

The app helped me involve in others' S\&P decisions  & 3.62 & 1.02 & -0.58 & -0.13 & 
4.17 & 0.92 & -1.10 & 0.94 & 598*** &  \textbf{<0.001} \\

The app helped us involve in whole community's S\&P decisions  & 3.75 & 1.00 & -0.58 & -0.13 & 
4.16 & 0.82 & -0.66 & -0.23 & 522*** &  \textbf{<0.001} \\

 \underline{\textbf{Community Participation:}}  & & & & & & & & & & \\ 
The app enabled me to participate in community's S\&P & 3.85 & 0.90 & -0.60 & 0.18 &
4.12 & 0.87 & -0.91 & 0.82 & 760* &  \textbf{0.021} \\

The app enabled others to participate in community's S\&P  & 3.87 & 0.84 & -0.52 & 0.42 & 
4.17 & 0.91 & -0.87 & -0.10 & 771.5* &  \textbf{0.011} \\

The app enabled whole community to participate in everyone's S\&P   & 3.85 & 0.90 & -0.69 & 0.32 & 
4.25 & 0.89 & -0.99 & 0.12 & 701** &  \textbf{0.003} \\ 

 \underline{\textbf{Community Trust:}}  & & & & & & & & & & \\ 

I trust others to protect my information  & 3.99 & 0.28 & -1.18 & 0.87 &
4.15 & 0.23 & -1.21 & 1.62 & 1012.5 &  0.166 \\

I trust others to give me advice  & 4.13 & 0.24 & -1.10 & 1.10 & 
4.23 & 0.20 & -1.23 & 2.44 & 967.5 &  0.356 \\

Others trust me to protect their information  & 4.04 & 0.24 & -0.81 & 0.10 & 
4.28 & 0.19 & -0.49 & -1.02 & 651* &  \textbf{0.016} \\

Others trust me to give them advice  & 3.96 & 0.25 & -0.72 & 0.11 & 
4.18 & 0.22 & -1.17 & 2.04 & 856* &  \textbf{0.023} \\
\bottomrule
*p\textless{}.05; **p\textless{}.01; ***p \textless .001  & & & & & & & & & & \\

\end{tabular}
\end{table*}



\end{appendices}
\end{document}